\documentclass[editedvolume,numreferences]{crckbked}
\usepackage[dvips]{graphicx}
\begin{document}

%\tableofcontents
\begin{article}
\begin{opening}

\title{Shot Noise of Cotunneling Current}
\runningtitle{Shot Noise of Cotunneling Current}

\author{Eugene \surname{Sukhorukov}}
%\runningauthor{Eugene \surname{Sukhorukov}}
\institute{D\'epartement de Physique Th\'eorique, Universit\'e de
Gen\`eve, CH-1211 Gen\`eve 4, Switzerland.}  

\author{Guido \surname{Burkard}}
\author{Daniel \surname{Loss}}
\institute{Department of Physics and Astronomy, University of Basel, 
Klingelbergstrasse 82, CH-4056 Basel, Switzerland}  

\begin{abstract}
We study the noise of the cotunneling current through one or several 
tunnel-coupled quantum dots in the Coulomb blockade regime. The various
regimes of weak and strong, elastic and inelastic cotunneling are
analyzed for quantum-dot systems (QDS) with few-level, nearly-degenerate, and
continuous electronic spectra. In the case of weak cotunneling we prove 
a non-equilibrium fluctuation-dissipation theorem which leads to 
a universal expression for the noise-to-current ratio (Fano factor).
The noise of strong inelastic cotunneling can be 
super-Poissonian due to switching between QDS states carrying currents 
of different strengths. The transport through a double-dot (DD) system 
shows an Aharonov-Bohm effect both in noise and current. 
In the case of cotunneling through a QDS with a continuous energy spectrum
the Fano factor is very close to one.

\end{abstract}           

\end{opening}

\section{Introduction}
\label{introduction}

\begin{figure}[htb]
\centerline{\includegraphics[width=16pc]{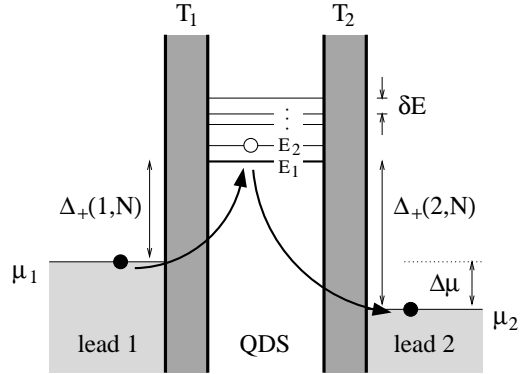}}
\caption{
The quantum dot system (QDS) is coupled to
two external leads $l=1,2$ via tunneling barriers.
The tunneling between the QDS and the leads
is parametrized by the tunneling amplitudes $T_l$,
where the lead and QDS quantum numbers $k$ and $p$ have
been dropped for simplicity, see Eq.~(\ref{tunneling}).
The leads are at the chemical potentials $\mu_{1,2}$, with an applied
bias $\Delta\mu=\mu_1 -\mu_2$.
The eigenstates of the QDS with one
added electron ($N+1$ electrons in total) are indicated by their energies
$E_1$, $E_2$, etc., with average level-spacing $\delta E$.
In the cotunneling regime there is
a finite energy cost
$\Delta_{\pm}(l,N)>0$
for the electron tunneling from the Fermi level of the lead $l$
to the QDS ($+$) and vice versa ($-$),
so that only processes of second order in $V$ (visualized by two arrows) 
are allowed.}
\label{energies}
\end{figure}

In recent years, there has been great interest in
the shot noise 
in mesoscopic
systems~\cite{review}, because it contains additional information 
about correlations, which is not contained, e.g., in
the linear response conductance. The shot noise is characterized 
by the Fano factor $F=S/eI$, the dimensionless ratio of the zero-frequency
noise power $S$ to the average current $I$.
While it assumes the Poissonian value $F=1$ in the absence of correlations,
it becomes suppressed or enhanced when correlations set in as e.g.
imposed by the Pauli principle or due to interaction effects.
In the present paper we study the shot noise of the 
cotunneling~\cite{averinazarov,Glattli} current.
We consider the transport through a quantum-dot system (QDS)
in the Coulomb blockade (CB) regime, in which the quantization of
charge on the QDS leads to a suppression of the sequential tunneling
current except under certain resonant conditions.  We consider the transport
away from these resonances and study the next-order contribution to
the current~\footnote{
The majority of papers on the noise of quantum dots consider
the sequential tunneling regime, where a classical description
(``orthodox'' theory) is applicable~\cite{AL}.
In this regime the noise is generally suppressed below its 
full Poissonian value $F=1$.
This suppression can be interpreted~\cite{SBL} 
as being a result of the natural correlations imposed 
by charge conservation.} (see Fig.~\ref{energies}). 
We find  that in the weak cotunneling 
regime, i.e. when the cotunneling rate $I/e$ is small compared
to the intrinsic relaxation rate $w_{\rm in}$ of the QDS to its equilibrium
state due to the coupling to the environment, $I/e\gg w_{\rm in}$,
the zero-frequency noise  takes on its Poissonian value, as first obtained 
for a special case in~\cite{LS}. This result is generalized here, and we find a 
universal relation between noise and current for the QDS
in the first nonvanishing order in the tunneling perturbation. 
Because of the universal character of this result Eq.~(\ref{DB-FDT})
we call it the nonequilibrium fluctuation-dissipation theorem 
(FDT)~\cite{Rogovin} in analogy with linear response theory.

One might expect however that the cotunneling,
being a two-particle process, may lead to strong correlations in the shot
noise and to the deviation of the Fano factor from its Poissonian value $F=1$. 
We show in Sec.~\ref{Super-Poissonian} that this is indeed the case 
for the regime of strong cotunneling, 
$I/e\gg w_{\rm in}$. Specifically, for a two-level QDS we predict giant (divergent) 
super-Poissonian noise \cite{RDB} (see Sec.~\ref{degenerate}): 
The QDS goes into an unstable mode where it switches
between states 1 and 2 with (generally) different currents.
In Sec.~\ref{DD-system} we consider the transport through a double-dot (DD) system 
as an example to illustrate this effect (see Eq.~(\ref{DD-noise})
and Fig.~\ref{double-d}). The Fano factor turns out to be
a periodic function of the magnetic flux through the DD leading to
an Aharonov-Bohm effect in the noise \cite{AB}. In the case of weak cotunneling 
we concentrate on the average current through the DD and find that 
it shows Aharonov-Bohm oscillations, which are a two-particle effect
sensitive to spin entanglement. 

Finally, in Sec.~\ref{continuum} 
we discuss the cotunneling through large QDS 
with a continuum spectrum.
In this case the correlations in the cotunneling current
described above do not play an essential role.  
In the regime of low bias, 
elastic cotunneling dominates transport,\cite{averinazarov}
and thus the noise is Poissonian.
In the opposite case of large bias, 
the transport is governed by inelastic cotunneling,
and in Sec.~\ref{continuum} we study heating effects which
are relevant in this regime.

\section{Model system}

In general, the QDS can contain several dots, which can be
coupled by tunnel junctions, the DD being a particular
example~\cite{LS}.
The QDS is assumed to be weakly coupled to external metallic leads which are
kept at equilibrium with their associated reservoirs at the chemical
potentials $\mu_l$, $l=1,2$,  where the  currents $I_l$ can be measured
and the average current $I$
through the QDS is defined by Eq.~(\ref{current-noise}).
Using a standard tunneling Hamiltonian approach~\cite{Mahan},
we write
\begin{eqnarray}
&& H=H_0+V\,,\quad H_0=H_L+H_S+H_{\rm int}\,, \label{Hamiltonian}   \\
&& H_L=\sum_{l=1,2}\sum_k\varepsilon_{k}c_{lk}^{\dag}c_{lk}\,, \quad
H_S=\sum_p\varepsilon_pd_p^{\dag}d_p\,, \label{QDS-leads} \\
&& V=\sum_{l=1,2}(D_l+D^{\dag}_l),\quad
D_l=\sum_{k,p}T_{lkp}c_{lk}^{\dag}d_p\,,\label{tunneling}
\end{eqnarray}
where  the terms $H_L$ and $H_S$  describe
the leads and QDS, respectively (with $k$ and $p$
from a complete set of quantum numbers),and
tunneling between leads  and  QDS is described
by the perturbation $V$. The interaction term $H_{\rm int}$
does not need to be specified for our proof of the universality 
of noise in Sec.~\ref{FDT-Double}.
The $N$-electron QDS is in the cotunneling regime where there is
a finite energy cost
$\Delta_{\pm}(l,N)>0$
for the electron tunneling from the Fermi level of the lead $l$
to the QDS ($+$) and vice versa ($-$). This energy cost is of the order 
of the charging energy $E_C$ and much larger than the temperature, 
$\Delta_{\pm}(l,N)\sim E_C\gg k_BT$,
so that only processes of second order in $V$ are allowed.

To describe the transport through the QDS we apply standard methods~\cite{Mahan}
and adiabatically switch on the perturbation $V$ in the distant past,
$t=t_0\to - \infty$. The perturbed state of the system
is described by the time-dependent density matrix
$\rho(t)=e^{-iH(t-t_0)}\rho_0 e^{iH(t-t_0)}$,
with $\rho_0$ being the grand canonical density matrix of the
unperturbed system, $\rho_0=Z^{-1}e^{-K/k_BT}$,
where we set $K=H_0-\sum_l\mu_lN_l$.
Because of tunneling the total number of electrons in each lead
$N_l=\sum_kc_{lk}^{\dag}c_{lk}$ is no longer conserved.
For the outgoing currents $\hat I_l=e\dot N_l$ we have
\begin{equation}
\hat I_l=ei\left[V,N_l\right] =ei(D^{\dag}_l-D_l)\,.
\label{currents}
\end{equation}
The observables of interest are the average current $I\equiv I_2=-I_1$
through the QDS,
and the spectral density of the noise
$S_{ll'}(\omega)=\int dt S_{ll'}(t)\exp(i\omega t)$,
\begin{equation}
I_l={\rm Tr}\rho(0)\hat I_l,\quad
S_{ll'}(t)={\rm Re}\,{\rm Tr}\,\rho(0)\delta I_l(t)\delta I_{l'}(0)\,,
\label{current-noise}
\end{equation}
where $\delta I_l=\hat I_l-I_l$.
Below we will use the interaction representation
where Eq.~(\ref{current-noise}) can be rewritten by replacing
$\rho(0)\to\rho_0$ and $\hat I_l(t)\to U^{\dag}(t)\hat I_l(t)U(t)$, with
\begin{equation}
U(t)=T\exp\left[-i\int^{t}_{-\infty}dt'\,V(t')\right]\,.
\label{U-Operator}
\end{equation}
In this representation, the time dependence of all operators is governed by the
unperturbed Hamiltonian $H_0$.

\section{Weak cotunneling: Non-equilibrium fluctuation-dissipation \\ theorem}
\label{FDT-Double}

In this section we prove the universality of noise of tunnel junctions
in the weak cotunneling regime $I/e\ll w_{\rm in}$ keeping the first
nonvanishing order in the tunneling Hamiltonian $V$.
Since our final result (\ref{DB-FDT}) can be applied to quite general
systems out-of-equilibrium
we call this result the non-equilibrium fluctuation-dissipation theorem (FDT).
In particular, the geometry of the QDS and the interaction $H_{\rm int}$ are 
completely arbitrary for the discussion of 
the non-equilibrium FDT in this section.

We note that the two currents $\hat I_l$ are not independent,
because $[\hat I_1,\hat I_2]\neq 0$, and thus all
correlators $S_{ll'}$ are nontrivial.
The charge accumulation on the QDS for a time of order
$\Delta_{\pm}^{-1}$ leads to an additional contribution to the noise
at finite frequency $\omega$. Thus, we expect
that for $\omega\sim\Delta_{\pm}$ the correlators
$S_{ll'}$ cannot be expressed through the steady-state
current $I$ only and thus $I$ has to be complemented by some
other dissipative counterparts, such as differential conductances
$G_{ll'}$. On the other hand, at low enough frequency, 
$\omega \ll \Delta_{\pm}$,
the charge conservation on the QDS requires
$\delta I_s=(\delta I_2+\delta I_1)/2\approx 0$.
Below we concentrate on the limit of low frequency
and neglect contributions of order of $\omega/\Delta_{\pm}$ to the noise power.
In  the Appendix we prove that $S_{ss}\sim (\omega/\Delta_{\pm})^2$ 
(see Eq.~(\ref{A09})),
and this allows us to redefine the current and the noise power as
$I\equiv I_d=(I_2-I_1)/2$ and 
$S(\omega)\equiv S_{dd}(\omega)$.~\footnote{
We note that charge fluctuations,
$\delta Q(t)\!=\!2\!\int_{-\infty}^{t}\!dt' \delta I_s(t')$,
on a QDS are also relevant for device
applications such as SET~\cite{problem}.
While we focus on current fluctuations in the present paper, we mention here
that in the cotunneling regime
the noise power $\langle \delta Q^2\rangle_{\omega}$
does not vanish at zero frequency,
$\langle \delta Q^2\rangle_{\omega=0}=
4\omega^{-2}S_{ss}(\omega)|_{\omega\to 0}\neq 0$.
Our formalism is also suitable for studying such charge fluctuations;
this will be addressed elsewhere.}
In addition we require that the QDS is in the cotunneling regime,
i.e. the temperature is low enough, $k_BT\ll \Delta_{\pm}$, although
the bias $\Delta\mu$ is arbitrary as soon as the sequential 
tunneling to the dot is forbidden, $\Delta_{\pm}>0$.
In this limit the current through a QDS arises due to the direct
hopping of an electron from one lead to another (through a virtual state
on the dot) with an amplitude which depends on the energy cost $\Delta_{\pm}$
of a virtual state. Although this process can change the state
of the QDS (inelastic cotunneling), the fast energy relaxation
in the weak cotunneling regime, $w_{\rm in}\gg I/e$, immediately
returns it to the equilibrium state (for the opposite case,
see Sec.~\ref{Super-Poissonian}). This allows
us to apply a perturbation expansion with respect to tunneling $V$
and to keep only first nonvanishing contributions, which we do next.

It is convenient to introduce the notation
$\bar D_l(t)\equiv\int_{-\infty}^{t}dt'\, D_l(t')$.
We notice that all relevant matrix elements,
$\langle N| D_l(t)|N+1\rangle\sim e^{-i\Delta_{+}t}$,
$\langle N-1| D_l(t)|N\rangle\sim e^{i\Delta_{-}t}$,
are fast oscillating functions of time. Thus, under the above
conditions we can write $\bar D_l(\infty)=0$,
and even more general, $\int_{-\infty}^{+\infty}dt\, D_l(t)e^{\pm i\omega t}=0$
(note that we have assumed earlier that $\omega\ll \Delta_\pm$).
Using these equalities and the cyclic property
of the trace we obtain the following results (for details of the derivation,
see Appendix A),
\begin{eqnarray}
&&I=e\int\limits_{-\infty}^{\infty}dt\,
\langle [A^{\dag}(t),A(0)]\rangle,\qquad
A=D_2\bar D^{\dag}_1+D^{\dag}_1\bar D_2\,,
\label{DB-current}\\
&&S(\omega)=e^2\int\limits^{\infty}_{-\infty}dt\,
\cos (\omega t)\langle\{A^{\dag}(t),A(0)\}\rangle\,,
\label{DB-noise}
\end{eqnarray}
where we have dropped a small contribution of order
$\omega/\Delta_{\pm}$ and used the notation 
$\langle\ldots\rangle={\rm Tr}\rho_0(\ldots)$.

Next we apply the spectral decomposition
to the correlators Eqs.~(\ref{DB-current}) and~(\ref{DB-noise}),
a similar procedure to that which also leads
to the equilibrium fluctuation-dissipation theorem.
The crucial observation is that $[H_0,N_l]=0$, $l=1,2$.
Therefore, we are allowed to use for our spectral decomposition the basis
$|{\bf n}\rangle=| E_{{\bf n}},N_1,N_2\rangle$ of eigenstates
of the operator $K=H_0-\sum_l\mu_l N_l$, which also diagonalizes the grand-canonical
density matrix $\rho_0$, $\rho_{{\bf n}}=\langle {\bf n}|\rho_0 |{\bf n} \rangle
=Z^{-1}\exp[-E_{{\bf n}}/k_BT]$.
We introduce the spectral function,
\begin{equation}
{\cal A}(\omega) =
2\pi \sum_{ {\bf n}, {\bf m}}(\rho_{{\bf n}} +\rho_{{\bf m}})
|\langle {\bf m}|A|{\bf n}\rangle|^2
\delta(\omega+E_{{\bf n}}-E_{{\bf m}})\,,
\label{spectral}
\end{equation}
and rewrite Eqs.~(\ref{DB-current}) and~(\ref{DB-noise})
in the matrix form in the basis $| {\bf n}\rangle$
taking into account that
the operator $A$, which plays the role of the effective cotunneling amplitude, 
creates (annihilates) an electron in the lead 2 (1) 
(see Eqs.~(\ref{tunneling}) and~(\ref{DB-current})). 
We obtain following expressions
\begin{eqnarray}
&&  I(\Delta\mu)
=e\tanh\left[\frac{\Delta\mu}{2k_BT}\right]{\cal A}(\Delta\mu)\,,
\label{DB-current2} \\
&& S(\omega,\Delta\mu)=\frac{e^2}{2}\sum_{\pm}
{\cal A}(\Delta\mu\pm\omega)\,.
\label{DB-noise2}
\end{eqnarray}
We note that because of additional integration over time $t$ in the 
amplitude $A$ (see Eq.~(\ref{DB-current})), the spectral density 
${\cal A}$ depends on $\mu_1$ and $\mu_2$ separately.
However, away from the resonances, $\omega\ll\Delta_\pm$,
only $\Delta\mu$-dependence is essential, and thus ${\cal A}$
can be regarded as being one-parameter 
function.~\footnote{To be more precise, we neglect small
$\omega$-shift of the energy denominators $\Delta_\pm$, which is equivalent
to neglecting small terms of order $\omega/\Delta_\pm$ in Eq.~(\ref{DB-noise2}).}
Comparing Eqs.~(\ref{DB-current2}) and~(\ref{DB-noise2}),
we obtain
\begin{equation}
 S(\omega,\Delta\mu)=\frac{e}{2}\sum_{\pm}
\coth\left[\frac{\Delta\mu\pm\omega}{2k_BT}\right]
I(\Delta\mu\pm\omega)
\label{DB-FDT}
\end{equation}
up to small terms on the order of $\omega/\Delta_\pm$. 
This equation represents our nonequilibrium FDT for the
transport through a QDS in the weak cotunneling regime.
A special case with $T, \omega=0$, giving $S=eI$, has
been derived earlier~\cite{LS}.
To conclude this section we would like to list again the conditions
used in the derivation.
The universality of noise to current relation
Eq.~(\ref{DB-FDT}) proven here is valid in the regime in which
it is sufficient to keep the  first nonvanishing order in the tunneling $V$
which contributes to transport and noise.
This means that  the QDS is in the weak cotunneling regime
with $\omega, k_BT \ll \Delta_{\pm }$, and
$I/e\ll w_{\rm in}$.

\section{Strong cotunneling: Correlation correction to noise}
\label{Super-Poissonian}

In this section we consider the QDS in the
strong cotunneling regime, $w_{\rm in}\ll I/e$.
Under this assumption the intrinsic relaxation in the QDS is very slow and will in fact
be neglected.  Thermal equilibration can only take place via coupling 
to the leads (see Sec.~\ref{continuum}).
Due to this slow relaxation in the QDS we find that
there are non-Poissonian correlations $\Delta S$ in the current through
the QDS because the QDS has a ``memory''; the state of the QDS after the
transmission of one electron influences the transmission of the next electron. 
The microscopic theory of strong cotunneling has been developed in Ref.~\cite{SBL}
based on the density-operator formalism and using the projection operator technique. 
Here we discuss the assumptions and present the results of the theory,
equations (\ref{MasterEq}), (\ref{golden-rule}), and
(\ref{AvCurrent}-\ref{correlator-delta-s}), 
which are the basis for our further
analysis in the Secs.\ \ref{degenerate} and \ref{DD-system}.

First, we assume that
the system and bath are coupled only weakly and only via the perturbation
$V$, Eq.~(\ref{tunneling}).  The interaction part $H_{\rm int}$ of the 
unperturbed Hamiltonian $H_0$, Eq.~(\ref{Hamiltonian}), must therefore
be separable into a QDS and a lead part,
$H_{\rm int} = H_S^{\rm int}+H_L^{\rm int}$.
Moreover, $H_0$ conserves the number of electrons in the leads,
$[H_0,N_l]=0$, where $N_l=\sum_{k} c_{lk}^\dagger c_{lk}$.
The assumption of weak coupling allows us to keep only the second-order in $V$
contributions to the ``golden rule'' rates 
(\ref{golden-rule}) for the Master equation (\ref{MasterEq}).

Second, we assume that in the distant past, $t_0\rightarrow -\infty$, 
the system is in an equilibrium state
\begin{equation}\label{initial}
  \rho _0 = \rho_{S}\otimes\rho_L,\quad%\quad\quad
  \rho_L = \frac{1}{Z_L}e^{-K_L/k_B T},
\end{equation}
where $Z_L={\rm Tr}\, \exp[-K_L/k_B T]$, $K_L=H_L-\sum_l\mu_l N_l$,
and $\mu_l$ is the chemical potential of lead $l$.
Note that both leads are kept at the same temperature $T$.
Physically, the product form of $\rho_0$ in Eq.~(\ref{initial}) describes the
absence of correlations between the QDS and the leads in the initial state at $t_0$.
Furthermore, we assume that the initial state $\rho_0$
is diagonal in the eigenbasis of $H_0$, i.e. that the initial state is
an incoherent mixture of eigenstates of the free Hamiltonian.

Finally, we consider the low-frequency noise, $\omega\ll\Delta_{\pm}$,
i.e.\ we neglect the accumulation of the charge on the QDS (in the same way
as in the Sec.~\ref{FDT-Double}). Thus we can write
$S_{ll}(\omega) = -S_{l\neq l'}(\omega)\equiv S(\omega)$.
This restriction will be lifted in the end of 
the Sec.~\ref{DD-weak}.

We note that the above assumptions limit the generality of the results of present
section as compared to those of Sec.~\ref{FDT-Double}. On the other hand,
they allow us to reduce the problem of the noise calculations to the solution 
of the Master equation
\begin{equation}
  \label{MasterEq}
  \dot\rho_n(t) = \sum_m\left[ w_{nm}\rho_m(t) - w_{mn}\rho_n(t)\right],
\end{equation}
with the stationary state condition
$\sum_m(w_{nm}\bar\rho_m - w_{mn}\bar\rho_n)=0$.
This ``classical'' master equation describes the dynamics of the QDS,
i.e.\ it describes the rates with which the probabilities $\rho_n$ for
the QDS being in state $|n\rangle$ change.
The rates
$w_{nm} = \sum_{l,l'=1,2} w_{nm}(l',l)$
are the sums of second-order ``golden rule'' rates
\begin{equation}
w_{nm}(l',l) = 2\pi \sum_{\bar{m},\bar{n}}
   |\langle{\bf n}|(D^\dagger_l, D_{l'})|{\bf m}\rangle|^2
\delta (E_{\bf m} - E_{\bf n}-\Delta\mu_{ll'})  \rho_{L,\bar{m}}.
\label{golden-rule}
\end{equation}
for all possible cotunneling transitions from lead $l$ to lead $l'$.
In the last expression, $\Delta\mu_{ll'} = \mu_l-\mu_{l'}$ denotes 
the chemical potential drop between lead $l$ and lead $l'$, and
$\rho_{L,\bar{m}}=\langle \bar{m}|\rho_L|\bar{m} \rangle $.
We have defined the second order hopping operator
\begin{equation}
(D^\dagger_l, D_{l'})= 
D_{l'} \bar{D}^\dagger_l + D^\dagger_l \bar{D}_{l'},
\label{matrix-element}
\end{equation}
where $D_l$ is given in Eq.~(\ref{tunneling}), and
$\bar{D_l}=\int_{-\infty}^0 D_l(t)dt$.
Note, that $(D^\dagger_l, D_{l'})$ is the amplitude of cotunneling 
from the lead $l$ to the lead $l'$ (in particular, we can write 
$A=(D^\dagger_1, D_2)$, see Eq.~(\ref{DB-current})). 
The combined index ${\bf m}=(m,\bar{m})$ contains both the QDS index
$m$ and the lead index $\bar{m}$. Correspondingly,
the basis states used above are $|{\bf m}\rangle = |m\rangle |\bar{m}\rangle$
with energy $E_{\bf m}=E_m+E_{\bar{m}}$,
where $|m\rangle$ is an eigenstate of $H_S+H_S^{\rm int}$ 
with energy $E_m$, and $|\bar{m}\rangle$ is an eigenstate of
$H_L+H_L^{\rm int}-\sum_l \mu_l N_l$ with energy $E_{\bar{m}}$.

For the average current $I$ and the noise power $S(\omega)$ we obtain
\cite{SBL}
\begin{eqnarray}
  I & = & e\sum_{mn}w^I_{nm}\bar\rho_m,\quad
  w^I_{nm} =w_{nm}(2,1) - w_{nm}(1,2),
\label{AvCurrent}\\
S(\omega) &=&  e^2\sum_{mn}[w_{nm}(2,1) + w_{nm}(1,2)]\bar\rho_m + \Delta S(\omega),
             \label{NDnoise}\\
  \Delta S(\omega)
    &=& e^2 \!\!\!\!\!\!\sum_{n,m,n',m'} w^I_{nm} \delta \rho_{mn'}(\omega) w^I_{n'm'}\bar\rho_{m'},\label{correlator-delta-s}
\end{eqnarray}
where $\delta\rho_{nm}(\omega) = \rho_{nm}(\omega)-2\pi\delta(\omega)\bar\rho_n$, and
$\bar\rho_n$ is the stationary density matrix.
Here, $\rho_{nm}(\omega)$ is the Fourier-transformed
conditional density matrix, which is obtained from the {\em symmetrized}
solution $\rho_n(t)=\rho_n(-t)$ of the
master equation Eq.~(\ref{MasterEq}) with the initial condition
$\rho_n(0)=\delta_{nm}$.

An explicit result
for the noise in this case can be obtained by making further
assumptions about the QDS and the coupling to the leads, see the
following sections.  For the general case, we only estimate $\Delta S$.
The current is of the order $I\sim ew$, with $w$ some typical
value of the cotunneling rate $w_{nm}$, and thus 
$\delta I\sim ew$. The time between switching from one dot-state
to another due to cotunneling is approximately $\tau\sim w^{-1}$.
The correction $\Delta S$ to the Poissonian noise can be estimated
as $\Delta S\sim\delta I^2\tau\sim e^2w$, which is of the same
order as the Poissonian contribution $e I\sim e^2 w$.
Thus the correction to the Fano factor is of order unity.
(Note however, that under certain conditions the Fano factor can diverge,
see Secs.\ \ref{degenerate} and \ref{DD-system}.)
In contrast to this, we find that for  elastic cotunneling
the off-diagonal rates vanish, $w_{nm}\propto \delta_{nm}$, and
therefore $\delta\rho_{nn}=0$ and $\Delta S=0$.
Moreover, at zero temperature, either $w_{nn}(2,1)$ or $w_{nn}(1,2)$ must
be zero (depending on the sign of the bias $\Delta\mu$). 
As a consequence, for elastic cotunneling we find Poissonian noise,
$F=S(0)/e|I|=1$.

\section{Cotunneling through nearly degenerate states}
\label{degenerate}

Suppose the QDS has nearly degenerate states
with energies $E_n$,  and
level spacing $\delta\! E_{nm}=E_n - E_m$,
which is much smaller
than the average level spacing $\delta E$. In the regime,
$\Delta\mu, k_BT, \delta\! E_{nm}\ll \delta E$, the only allowed
cotunneling processes
are the transitions between nearly degenerate states.
The noise power is given by Eqs.\ (\ref{NDnoise}) and
(\ref{correlator-delta-s}),
and below we calculate the correlation correction to the noise, $\Delta S$.
To proceed with our calculation
we rewrite Eq.~(\ref{MasterEq}) for $\delta\rho(t)$
as a second-order differential equation in matrix form
\begin{equation}
\delta\ddot\rho(t) = W^2\delta\rho(t),\quad \delta\rho(0)=1-\bar\rho ,
\label{second-eq}
\end{equation}
where $W$ is defined as
$W_{nm} = w_{nm} - \delta_{nm}\sum_{m'}w_{m'n}$.
We solve this equation by
Fourier transformation,
\begin{equation}
\delta\rho(\omega)=-\frac{2W}{W^2+\omega^2 1},
\label{delta-rho}
\end{equation}
where we have used $W\bar\rho=0$.
We substitute $\delta\rho$ from this equation into
Eq.~(\ref{correlator-delta-s})
and write the result in a compact matrix form,
\begin{equation}
\Delta S(\omega)=
-e^2\sum_{n,m}\left[w^{I}\frac{2W}{W^2+\omega^2
1}w^{I}\bar\rho\right]_{nm}.
\label{result01}
\end{equation}
This equation gives the formal solution of the noise problem for
nearly degenerate states. As an example we consider a two-level system.

Using the detailed balance equation, $w_{21}\rho_1=w_{12}\rho_2$, we obtain
for
the stationary probabilities $\rho_1=w_{12}/(w_{12}+w_{21})$, and
$\rho_2=w_{21}/(w_{12}+w_{21})$. From Eq.~(\ref{AvCurrent}) we get
\begin{equation}
I=e\frac{w_{12}(w^{I}_{11}+w^{I}_{21})+w_{21}(w^{I}_{22}+w^{I}_{12})}
{w_{12}+w_{21}}.
\label{two-current}
\end{equation}
A straightforward calculation with the help of Eq.~(\ref{delta-rho})
gives for the correction to the Poissonian noise
\begin{eqnarray}
\Delta S(\omega) & = &
\frac{2e^2(w^{I}_{11}+w^{I}_{21}-w^{I}_{22}-w^{I}_{12})}
{(w_{12}+w_{21})[\omega^2+(w_{12}+w_{21})^2]}\times\nonumber \\
& \times  & \left[w^{I}_{11}w_{12}w_{21}+w^{I}_{12}w^2_{21}-
(1\leftrightarrow 2)\right].
\label{two-noise}
\end{eqnarray}
In particular, the zero frequency noise
$\Delta S(0)$
diverges if the ``off-diagonal'' rates $w_{nm}$ vanish.
This divergence has to be cut at
$\omega$, or at the relaxation rate $w_{\rm in}$ due to coupling to
the bath (since $w_{12}$ in this case
has to be replaced with $w_{12}+w_{\rm in}$).
The physical origin of the divergence is rather
transparent: If the off-diagonal rates $w_{12},w_{21}$
are small, the QDS goes into an
unstable state where it switches between states 1 and 2 with
different currents in general~\cite{comment5}.
The longer the QDS stays in the
state 1 or 2 the larger the zero-frequency noise power is. However,
if $w^{I}_{11}+w^{I}_{21}=w^{I}_{22}+w^{I}_{12}$,
then $\Delta S(\omega)$ is suppressed to 0. For
instance, for the QDS in the spin-degenerate state with an odd
number of electrons $\Delta S(\omega)=0$, since the two states
$|\uparrow\rangle$ and $|\downarrow\rangle$ are physically
equivalent. The other example of such a suppression
of the correlation correction $\Delta S$
to noise is given by a multi-level QDS, $\delta E\ll E_C$,
where the off-diagonal rates are
small compared to the diagonal (elastic) rates~\cite{averinazarov}.
Indeed, since the main contribution to the elastic rates comes
from transitions through many virtual states,
which do not participate in inelastic cotunneling,
they do not depend on the initial
conditions, $w^I_{11}=w^I_{22}$, and cancel in the numerator of
Eq.~(\ref{two-noise}), while they are still present in the current.
Thus the correction $\Delta S/I$ vanishes in this case.
Further below in this section we consider a few-level QDS,
$\delta E\sim E_C$, where $\Delta S\neq 0$.

To simplify further analysis we consider for a moment the case, where
the singularity in the noise is most pronounced, namely,
$\omega=0$ and $|\delta\! E_{12}|\ll\Delta\mu, k_BT$, so that
$w^I_{12}=w^I_{21}$, and $w_{12}=w_{21}$.
Then, from Eqs.\ (\ref{two-current}) and (\ref{two-noise})
we obtain
\begin{eqnarray}
&& I=\frac{1}{2}(I_{1}+I_{2})\,,\quad I_{n}=e\sum_{m=1,2}w^I_{mn}\,,
\label{two-current2} \\
&& \Delta S(0)=\frac{(I_{1}-I_{2})^2}{4w_{12}}\,,
\label{two-noise2}
\end{eqnarray}
where $I_n$ is the current through the $n$-th level of the QDS.
Thus in case $|\delta\! E_{12}|\ll\Delta\mu, k_BT$ the
following regimes have to be distinguished:
(1) If $k_BT\lesssim\Delta\mu$, then $I_n\propto\Delta\mu$,
$w_{12}\propto\Delta\mu$, and thus both, the total current
$I=e^{-1}G_D\Delta\mu$, and
the total noise $S=FG_D\Delta\mu$ are linear in the bias $\Delta\mu$
(here $G_D$ is the conductance of the QDS).
The total shot noise in this regime is super-Poissonian with the Fano factor
$F\sim I/(ew_{12})\gg 1$.
(2) In the regime $\Delta\mu\lesssim k_BT\lesssim F^{1/2}\Delta\mu$
the noise correction (\ref{two-noise2}) arises because of the thermal
switching
the QDS between two states $n=1,2$, where the currents are linear in the
bias,
$I_n\sim G_D\Delta\mu/e$. The rate of switching is $w_{12}\propto k_BT$, and
thus
$\Delta S\sim FG_D\Delta\mu^2/(k_BT)$. Since $k_BT/\Delta\mu\lesssim
F^{1/2}$, the
noise correction $\Delta S$
is the dominant contribution to the noise, and thus the total noise $S$
can be interpreted as being a thermal telegraph noise~\cite{Kogan}.
(3) Finally, in the regime $F^{1/2}\Delta\mu\lesssim k_BT$
the first term on the rhs of
Eq.\ (\ref{NDnoise}) is the dominant contribution, and the total noise
becomes
an equilibrium Nyquist noise, $S=2G_Dk_BT$.

\begin{figure}
\centerline{\includegraphics[width=26pc]{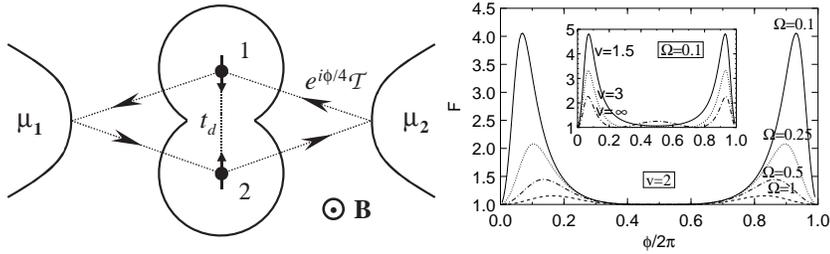}}
\caption{
Left:
Double-dot (DD) system containing two electrons and
being weakly coupled to metallic leads 1, 2,
each of which is at the chemical potential
$\mu_1$, $\mu_2$. The tunneling amplitudes between dots and leads are
denoted
by $\cal T$. The tunneling ($t_d$) between the dots
results in a singlet-triplet splitting $J\sim t_d^2/U$
with the singlet being a ground state.
The tunneling path between dots
and leads 1 and 2 forms a closed loop (shown by arrows)
so that the Aharonov-Bohm phase $\phi$ will be accumulated by
an electron traversing the DD.
Right:
The Fano factor $F=S(\omega)/I$, with the noise power
$S(\omega)$ given in Eqs.~(\ref{NDnoise}) and
(\ref{DD-noise}), is
plotted as a function of the Aharonov-Bohm phase $\phi$
for the normalized bias $v\equiv \Delta\mu/J=2$ and for four
different normalized frequencies
$\Omega\equiv\omega /[G(2\Delta\mu - J)]=0.1$, $0.25$,
$0.5$, and $1$. Inset: the same, but with fixed frequency
$\Omega=0.1$, where the bias $v$ takes the values
$1.5$, $3$, and $\infty$.
}
\label{double-d}
\end{figure}

\section{Noise of double-dot system: 
Two-particle Aharonov-Bohm effect}
\label{DD-system}

We notice that
for the noise power to be divergent the off-diagonal rates
$w_{12}$ and $w_{21}$ have to vanish simultaneously.
However, the matrix $w_{nm}$ is not symmetric since the off-diagonal
rates depend on the bias in a different way. On the other hand,
both rates contain the same matrix element of the cotunneling amplitude
$(D^{\dag}_l,D_{l'})$, see Eqs.~(\ref{golden-rule}) and
(\ref{matrix-element}).
Although in general this matrix element is not small, it can vanish
because of different symmetries of the two states. To illustrate this effect
we consider the transport through a double-dot (DD) system
(see Ref.~\cite{LS} for details) as an example. Two leads
are equally coupled to two dots  in such a way
that a closed loop is formed, and the dots are also connected,
see Fig.~\ref{double-d}.
Thus, in a magnetic field the tunneling is described by the Hamiltonian
Eq.~(\ref{tunneling}) with
\begin{eqnarray}
&&D_l=\sum_{s,j}T_{lj}c^{\dag}_{ls}d_{js}\,,
\qquad l,j=1,2\,,
\label{DD-D}\\
&&T_{11}=T_{22}=T^{*}_{12}=T^{*}_{21}=
e^{i\phi/4}{\cal T}\,,
\label{equal}
\end{eqnarray}
where the last equation expresses the equal coupling of dots and leads
and $\phi$ is the Aharonov-Bohm phase.
Each dot contains one electron, and weak tunneling $t_d$ between the dots
causes
the exchange splitting~\cite{Burkard}
$J\sim t_d^2/U$ (with $U$ being the on-site repulsion)
between one spin singlet and three triplets
\begin{eqnarray}
&&|S\rangle=\frac{1}{\sqrt{2}}
[d^{\dag}_{1\uparrow}d^{\dag}_{2\downarrow}\!-
d^{\dag}_{1\downarrow}d^{\dag}_{2\uparrow}]|0\rangle\,,
\nonumber\\
&&|T_0\rangle=\frac{1}{\sqrt{2}}
[d^{\dag}_{1\uparrow}d^{\dag}_{2\downarrow}\!+
d^{\dag}_{1\downarrow}d^{\dag}_{2\uparrow}]|0\rangle\,,
\label{basis}\\
&&|T_+\rangle=d^{\dag}_{1\uparrow}d^{\dag}_{2\uparrow}|0\rangle\,,
\quad
|T_-\rangle=d^{\dag}_{1\downarrow}d^{\dag}_{2\downarrow}|0\rangle\,.
\nonumber
\end{eqnarray}
In the case of zero magnetic field, $\phi=0$, the tunneling Hamiltonian $V$
is symmetric with respect to the exchange of electrons, $1\leftrightarrow
2$.
Thus the matrix element of the cotunneling transition between the singlet
and three triplets $\langle S|V(E-H_0)^{-1}V|T_i\rangle$, $i=0,\pm$,
vanishes
because these states have different orbital symmetries.
A weak magnetic field breaks the symmetry, contributes to the off-diagonal
rates, and thereby reduces noise. Next, we consider weak and strong cotunneling
regimes.

\subsection{Weak cotunneling}
\label{DD-weak}

In this regime, $I/e\ll w_{\rm in}$, according to the non-equilibrium FDT
(see Sec.~\ref{FDT-Double}) the zero-frequency noise contains the same information
as the average current (the Fano factor $F=1$). 
Therefore, we first concentrate on current.
We focus on the  regime, $\Delta\mu\gg J$, where inelastic
cotunneling~\cite{inelastic} occurs with singlet and triplet contributions being
different, and where we can neglect the
dynamics generated by $J$ compared to the one generated by the bias ("slow
spins"). Close to the sequential tunneling
peak, $\Delta_{-}\ll\Delta_{+}\sim U$, we keep only the term  
$D^{\dag}_{1}\bar D_{2}$ in the amplitude (\ref{DB-current}).
After some calculations we obtain
\begin{eqnarray}\label{current4}
&&I=e^{-1}C(\varphi)G\Delta\mu ,\\
&&C(\varphi )=\sum_{s,s'}\left[\langle
d^{\dag}_{1s^{\prime}}d_{1s}d^{\dag}_{1s}d_{1s^{\prime}}\rangle +\cos
\varphi\langle d^{\dag}_{1s^{\prime}}d_{1s}d^{\dag}_{2s}d_{2s^{\prime}}
\rangle \right],\label{factor1}
\end{eqnarray}
where $G=\pi (e\nu{\cal T}^2/\Delta_{-})^2$ is the conductance
of a single dot in the cotunneling regime~\cite{Recher}, and
we assumed Fermi liquid leads with the tunneling
density of states $\nu$. Eq.\ (\ref{current4})
shows that the cotunneling
current depends on the properties of the equilibrium state of the DD
through
the coherence factor $C(\varphi )$ given in  (\ref{factor1}).
The first term in $C$ is the contribution
from the topologically trivial tunneling path (phase-incoherent part)
which runs from
lead 1 through, say, dot 1 to lead 2 and back. The second
term (phase-coherent part) in $C$ results from an exchange process of
electron
1 with electron 2 via the leads 1 and 2 such that a closed loop is formed
enclosing an area $A$ (see Fig.~\ref{double-d}).
Note that for singlet and triplets the initial and final spin states
are the same  after such an exchange process. Thus, in the presence of a
magnetic
field $B$, an Aharonov-Bohm phase factor  $\varphi=ABe/h$ is acquired.

Next, we evaluate $C(\varphi)$ explicitly
in the singlet-triplet basis (\ref{basis})
and discuss the applications to the physics of quantum entanglement
(see the Ref.~\cite{LS}). Note that only the singlet
$|S\rangle$ and the triplet $|T_0\rangle$ are entangled EPR pairs while
the remaining triplets are not (they factorize).
Assuming that the DD is in one of these states we obtain the
important
result
\begin{equation}\label{factor2}
C(\varphi )= 2\mp\cos \varphi \,\, .
\end{equation}
Thus, we see that the singlet (upper sign) and the triplets (lower sign)
contribute with opposite sign to the phase-coherent part of the
current.
One has to distinguish, however,
carefully the entangled from the non-entangled states.
The   phase-coherent part of the entangled states is
a genuine {\it two-particle} effect, while the one of the product states
cannot be distinguished from a phase-coherent {\it single-particle}
effect.
Indeed, this follows from the observation that the
phase-coherent part in $C$ factorizes for the product states
$T_{\pm}$ while it does not so for $S, T_0$. Also, for states such as
$|\uparrow\downarrow\rangle$ the coherent part of $C$ vanishes, showing that
two
different (and fixed) spin states cannot lead to a phase-coherent
contribution since we {\it know} which electron goes which part of the loop.

Finally, we present our results \cite{LS} for the high-frequency noise
in the quantum range of frequancies, $\omega\sim\Delta_{-}\ll\Delta_{+}$,
and in the slow-spin regime $\Delta\mu\gg J$. This range of frequancies is
beyond the regime of the applicability of the non-equilibrium FDT, and therefore
there is no simple relation between the average current and the noise 
(see the Sec.~\ref{FDT-Double}).  After
lengthy calculations using the perturbation expansion of (\ref{U-Operator})
up to third order in $V$
we obtain
\begin{eqnarray}\label{noise2}
& & S(\omega)=(e\pi\nu{\cal T}^2)^2
\left[X_{\omega}+X^{*}_{-\omega}\right],
\nonumber \\
& & ImX_{\omega}={{C(\varphi )}\over {2\omega}}\left[\theta (\mu_1-\omega
)-\theta (\mu_2-\omega )\right], \\
& & ReX_{\omega}={{C(\varphi )}\over {2\pi\omega}}sign(\mu_1-\mu_2+\omega
)\ln|{{(\mu_1+\omega )(\mu_2-\omega )}\over {\mu_1\mu_2}}|
\nonumber \\
& & -{1\over {2\pi \omega }}\left[\theta (\omega -\mu_1)\ln
|{{\mu_2-\omega }\over {\mu_2}}|+
\theta (\omega -\mu_2)\ln
|{{\mu_1-\omega}\over {\mu_1}}|\right],
\label{noise3}
\end{eqnarray}
where $\mu_{l}=\Delta_{-}(l)$. Thus the real part of $S(\omega)$ 
is even in $\omega$, while the imaginary part
is odd. A remarkable feature here is that the noise  acquires an imaginary 
(i.e. odd-frequency) part
for finite frequencies,
in contrast to single-barrier junctions, where Im$ S(\omega)$ always
vanishes
since we have $\delta I_1=-\delta I_2$ for all times.
In double-barrier junctions considered here we find that
at small enough bias
$\Delta\mu\ll\Delta_{-} =\left(\Delta_{-}(1)+\Delta_{-}(2)\right)/2$, the odd part,
Im$ S(\omega)$, given in (\ref{noise2})
exhibits  two narrow peaks at $\omega =\pm\mu$, which in real time lead to
slowly decaying oscillations,
\begin{equation}
S_{odd}(t)=
\pi(e\nu{\cal T}^2)^2C(\varphi )\frac{\sin(\Delta\mu t/2)}{t\Delta_{-}}\sin (t\Delta_{-}).
\end{equation}
These oscillations again depend on the phase-coherence factor $C$ with the
same
properties as discussed before. These oscillations can be interpreted as
a temporary build-up  of a charge-imbalance on
the DD during an uncertainty time $\sim 1/\Delta_{-}$, which results from
cotunneling of electrons and an associated time delay between  out- and
ingoing currents.

%\begin{figure}
%\centerline{\includegraphics[width=16pc]{pic3.eps}}
%\parbox{80mm}{\caption{
%The Fano factor $F=S(\omega)/I$, with the noise power
%$S(\omega)$ given in Eqs.~(\ref{NDnoise}) and
%(\ref{DD-noise}), and with the current through the DD, $I$,
%given in Eqs.~(\ref{two-current}) and (\ref{DD-rates}), is
%plotted as a function of the Aharonov-Bohm phase $\phi$
%for the normalized bias $v\equiv \Delta\mu/J=2$ and for four
%different normalized frequencies
%$\Omega\equiv\omega /[G(2\Delta\mu - J)]=0.1$, $0.25$,
%$0.5$, and $1$. Inset: the same, but with fixed frequency
%$\Omega=0.1$, where the bias $v$ takes the values
%$1.5$, $3$, and $\infty$.
%}}
%\label{noise-corr}
%\end{figure}

\subsection{Strong cotunneling}
\label{DD-strong}

The fact that in the perturbation $V$ all spin indices are traced out
helps us to map the four-level system to only two states
$|S\rangle$ and $|T\rangle$ classified according to the orbital
symmetry (since all triplets are antisymmetric in orbital space).
In Appendix B we derive the mapping to a two-level system
and calculate the transition rates $w_{nm}(l',l)$
($n,m=1$ for a singlet
and $n,m=2$ for all triplets)
using Eqs.~(\ref{golden-rule}) and (\ref{matrix-element})
with the operators $D_l$ given by Eq.~(\ref{DD-D}).
Doing this we obtain the following result
\begin{eqnarray}
&& w_{nm}(1,1)= w_{nm}(2,2)= w_{nm}(1,2)=0,\nonumber \\
&& w_{nm}=w^I_{nm}=
\frac{\pi}{2}\left(\frac{\nu{\cal T}^2}{\Delta_{-}}\right)^2 \nonumber \\
&&
\times\left\{\begin{array}{ll}
(1\!+\!\cos\phi)\Delta\mu & (1\!-\!\cos\phi)(\Delta\mu\!+\!J) \\
3(1\!-\!\cos\phi)(\Delta\mu\!-\!J) & 3(1\!+\!\cos\phi)\Delta\mu \\
\end{array}
\right\},
\label{DD-rates}
\end{eqnarray}
which holds close to the sequential tunneling peak,
$\Delta_{-}\ll\Delta_{+}\sim U$
(but still $\Delta_{-}\gg J,\Delta\mu$),
and for $\Delta\mu>J$. We substitute this equation into the
Eq.~(\ref{two-noise})
and write the correction $\Delta S(\omega)$ to the Poissonian noise
as a function of normalized bias $v=\Delta\mu/J$ and normalized
frequency $\Omega=e\omega /[G(2\Delta\mu - J)]$
\begin{equation}
\Delta S(\omega)= 6eGJ
\frac{(v^2-1)[1+(v-1)\cos\phi]^2(1-\cos\phi)}{(2v-1)^3[\Omega^2+(1-\cos\phi)
^2]}.
\label{DD-noise}
\end{equation}
From this equation it follows that
the noise power has singularities as a function of $\omega$ for zero
magnetic field,
and it has singularities at $\phi=2\pi m$ (where $m$ is integer) as a
function
of the magnetic field (see Fig.~\ref{double-d}).
We would like to emphasize that the noise is singular even if the exchange
between
the dots is weak, $J\ll \Delta\mu$. 
In the case $\Delta\mu<J$ the transition from the singlet
to the triplet is forbidden by conservation of energy, $w_{21}(2,1)=0$,
and we immediately obtain from Eq.~(\ref{two-noise}) that $\Delta
S(\omega)=0$,
i.e.\ the total noise is Poissonian (as it is always the case for elastic
cotunneling).
In the case of large bias, $\Delta\mu\gg J$,
two dots contribute independently to the current $I=2e^{-1}G\Delta\mu$,
and from Eq.\ (\ref{DD-noise}) we obtain the Fano factor
\begin{equation}
F=\frac{3}{8}\,\frac{\cos^2\!\phi(1-\cos\phi)}{\Omega^2+(1-\cos\phi)^2}+1,
\quad \Delta\mu\gg J.
\label{Fano}
\end{equation}
This Fano factor controls the transition to the telegraph noise and then to
the equilibrium noise at high temperature, as described above.
We notice that if the coupling of the dots to the leads is not equal, then
$w_{nm}(l,l)\neq 0$ serves as a cut-off of the singularity in
$\Delta S(\omega)$.

Finally, we remark that the Fano factor is a periodic
function of the phase $\phi$ (see Fig.~\ref{double-d});
this is nothing but an Aharonov-Bohm effect in the
noise of the cotunneling transport through the DD.
However, in contrast to the  Aharonov-Bohm effect
in the cotunneling current through the DD which has
been discussed earlier in the Sec.~\ref{DD-weak}, the noise
effect does not allow us to probe the ground state of the DD,
since the DD is already in a mixture of the singlet and three
triplet states.

\section{Cotunneling through continuum of single-electron states}
\label{continuum}

We consider now the transport through a multi-level QDS with 
$\delta E\ll E_C$. In the low bias regime, 
$\Delta\mu\ll (\delta E\, E_C)^{1/2}$,
the elastic cotunneling dominates transport~\cite{averinazarov},
and according to the results of Sec.~\ref{Super-Poissonian} 
the noise
is Poissonian. Here we consider the opposite regime of
inelastic cotunneling,
$\Delta\mu\gg (\delta E\, E_C)^{1/2}$.
Since a large number $M$ of levels participate in transport,
we can neglect the correlations which we have studied in  
Secs.\ \ref{degenerate} and \ref{DD-system}, since they become a $1/M$-effect. 
Instead, we concentrate on the heating effect, which
is not relevant for the 2-level system considered before. 
The condition for strong cotunneling has to be rewritten
in a single-particle form, $\tau_{\rm in}\gg\tau_c$, where
$\tau_{\rm in}$ is the single-particle energy relaxation 
time on the QDS due to the coupling
to the environment, and $\tau_c$ is the time of the cotunneling
transition, which can be estimated as $\tau_c\sim e\nu_D\Delta\mu/I$
(where $\nu_D$ is the density of QDS states).
Since the
energy relaxation rate on the QDS is small, 
the multiple cotunneling transitions can cause
high energy excitations on the dot, 
and this leads to a nonvanishing backward
tunneling, $w_{nm}(1,2)\neq 0$. 
In the absence of correlations between cotunneling events, 
Eqs.~(\ref{AvCurrent}) and (\ref{NDnoise}) can be rewritten in terms of
forward and backward tunneling currents $I_{+}$ and $I_{-}$,
\begin{eqnarray}
&& I=I_{+}-I_{-}\,,\quad S=e(I_{+}+I_{-}),
\label{IS} \\
&&I_{+}=e\sum_{n,m}w_{nm}(2,1)\bar\rho_m\,,\quad
I_{-}=e\sum_{n,m}w_{nm}(1,2)\bar\rho_m\,, \label{Ipm1}
\end{eqnarray}
where the transition rates are given by (\ref{golden-rule}).

It is convenient to rewrite the currents $I_{\pm}$ in a single-particle basis. 
To do so we substitute the rates Eq.~(\ref{golden-rule}) into
Eq.~(\ref{Ipm1}) and neglect the dependence of the
tunneling amplitudes Eq.~(\ref{tunneling}) on the quantum numbers
$k$ and $p$, $T_{lkp}\equiv T_{l}$, which is a reasonable assumption for QDS with
a large number of electrons. Then we define the distribution function on
the QDS as 
\begin{equation}
f(\varepsilon)=\nu_D^{-1} \sum_{p}\delta(\varepsilon -
\varepsilon_p){\rm Tr}\,\bar\rho d^{\dag}_{p}d_{p}
\end{equation}
and replace the summation
over $p$ with an integration over $\varepsilon$. 
Doing this we obtain the following expressions for $T=0$
\begin{eqnarray}
&& I_{\pm}  =  C_{\pm}\frac{G_1G_2}{2\pi e^3}
\left(\frac{1}{\Delta_{+}}+\frac{1}{\Delta_{-}}\right)^2(\Delta\mu)^3,
\label{Ipm2} \\
&& C_{\pm} = \frac{1}{\Delta\mu^3}
\int\!\!\int\! d\varepsilon d\varepsilon' 
\Theta(\varepsilon\! -\varepsilon'\!\pm\Delta\mu)
f(\varepsilon)[1-f(\varepsilon')],
\label{Cpm}
\end{eqnarray}
where $G_{1,2}=\pi e^2\nu\nu_D|T_{1,2}|^2$ 
are the tunneling conductances of
the barriers 1 and 2,
and where we have introduced the function 
$\Theta(\varepsilon)=\varepsilon\theta(\varepsilon)$
with $\theta(\varepsilon)$ being the step-function.
In particular, using the property
$\Theta(\varepsilon+\Delta\mu)-\Theta(\varepsilon-\Delta\mu)
=\varepsilon +\Delta\mu$
and fixing
\begin{equation}
\int d\varepsilon
[f(\varepsilon)-\theta({-\varepsilon})]=0,
\label{symmetry}
\end{equation}
(since $I_{\pm}$ given by
Eq.~(\ref{Ipm2}) and Eq.~(\ref{Cpm}) do not depend on the shift
$\varepsilon\to\varepsilon+const$) we arrive at the following
general expression for the cotunneling current
\begin{eqnarray}
&&I= \Lambda\,\frac{G_1G_2}{12\pi e^3}
\left(\frac{1}{\Delta_{+}}+\frac{1}{\Delta_{-}}\right)^2(\Delta\mu)^3,
\label{I-continuum}\\
&&\Lambda=1+12\Upsilon/(\Delta\mu)^2, \label{prefactor}\\
&&\Upsilon=\int d\varepsilon \varepsilon [f(\varepsilon)-
\theta(-\varepsilon)]\geq 0, \label{Upsilon}
\end{eqnarray}
where the value $\nu_D\Upsilon$ has the physical meaning of
the energy acquired by the QDS due to the cotunneling
current through it.

We have deliberately  introduced the functions $C_{\pm}$ in the
Eq.~(\ref{Ipm2}) to emphasize the fact that if the distribution
$f(\varepsilon)$ scales with the bias $\Delta\mu$ (i.e.\ $f$ is a
function of $\varepsilon/\Delta\mu$), then $C_{\pm}$ become
dimensionless universal numbers. Thus both, the prefactor $\Lambda$
(given by Eq.~(\ref{prefactor})) in the cotunneling current, and the
Fano factor,
\begin{equation}
F=\frac{C_{+}+C_{-}}{C_{+}-C_{-}},
\label{F-continuum}
\end{equation}
take their universal values, which do not depend on the bias
$\Delta\mu$. We consider now such universal regimes.
The first example is the case of weak cotunneling, 
$\tau_{\rm in}\ll\tau_c$, when the QDS is in its ground state,
$f(\varepsilon)=\theta (-\varepsilon)$, and the thermal energy of
the QDS vanishes, $\Upsilon=0$. Then $\Lambda=1$, and 
Eq.~(\ref{I-continuum}) reproduces the results of Ref.~\cite{averinazarov}. 
As we have already mentioned, the
backward current vanishes, $I_{-}=0$, and the Fano factor acquires
its full Poissonian value $F=1$, in agreement with our 
nonequilibrium FDT proven in Sec.\ \ref{FDT-Double}. 
In the limit of strong cotunneling,
$\tau_{\rm in}\gg\tau_c$, the energy relaxation 
on the QDS can be neglected.
Depending on the electron-electron scattering time $\tau_{ee}$ two
cases have to be distinguished: The regime of cold electrons
$\tau_{ee}\gg\tau_c$ and regime of hot electrons
$\tau_{ee}\ll\tau_c$ on the QDS. Below we discuss both
regimes in detail and demonstrate their universality.

\subsection{Cold electrons}
\label{cold}

In this regime  the electron-electron scattering on the QDS can be
neglected and the distribution $f(\varepsilon)$ has to be found
from the master equation Eq.~(\ref{MasterEq}). We multiply this
equation by $\nu_D^{-1}\sum_{p}\delta(\varepsilon -
\varepsilon_p) \langle n|d^{\dag}_{p}d_{p}|n\rangle$, sum over
$n$ and use the tunneling rates from Eq.~(\ref{golden-rule}). Doing this we
obtain the standard stationary kinetic equation which can be
written in the following form
\begin{eqnarray}
&& \int d\varepsilon'\sigma(\varepsilon'
-\varepsilon)f(\varepsilon')[1-f(\varepsilon)]   \nonumber \\
&& \qquad\qquad\qquad =\int d\varepsilon' \sigma(\varepsilon
-\varepsilon')f(\varepsilon)[1-f(\varepsilon')],
\label{kineq1}\\
&& \sigma(\varepsilon)=2\lambda\Theta(\varepsilon)+
\sum_{\pm}\Theta(\varepsilon \pm\Delta\mu),
\label{kernel}
\end{eqnarray}
where $\lambda=(G_1^2+G_2^2)/(2G_1G_2)\geq1$ arises from the equilibration
rates $w_{mn}(l,l)$.
(We assume that if the limits of the integration over energy $\varepsilon$ are not
specified, then the integral goes from $-\infty$ to $+\infty$.) From the form of this
equation we immediately conclude that its solution is a function
of $\varepsilon/\Delta\mu$, and thus the cold electron regime is universal
as defined in the previous section.
It is easy to check that the detailed balance does not hold, and in addition
$\sigma(\varepsilon)\neq\sigma(-\varepsilon)$. Thus we face a difficult problem
of solving Eq.~(\ref{kineq1}) in its full nonlinear form.
Fortunately, there is a way to avoid this problem and to reduce the equation
to a linear form which we show next.

We group all nonlinear terms on the rhs of Eq.~(\ref{kineq1}):
$\int
d\varepsilon'\sigma(\varepsilon'-\varepsilon)f(\varepsilon')=
h(\varepsilon)f(\varepsilon)$, where $h(\varepsilon)=\int
d\varepsilon'
\left\{\sigma(\varepsilon'-\varepsilon)f(\varepsilon')+
\sigma(\varepsilon-\varepsilon')[1-f(\varepsilon')]\right\}$. The
trick is to rewrite the function $h(\varepsilon)$ in terms of
known functions. For doing this we split the integral in
$h(\varepsilon)$ into two integrals over $\varepsilon'>0$ and
$\varepsilon'<0$, and then use Eq.~(\ref{symmetry}) and the
property of the kernel $\sigma(\varepsilon)-\sigma(-\varepsilon)=
2(1+\lambda)\varepsilon$ to regroup terms in such a way that
$h(\varepsilon)$ does not contain $f(\varepsilon)$ explicitly.
Taking into account Eq.~(\ref{Upsilon}) we
arrive at the following linear integral equation
\begin{equation}
\int d\varepsilon'\sigma(\varepsilon'-\varepsilon)f(\varepsilon')
=[(1+\lambda)(\varepsilon^2+2\Upsilon)+(\Delta\mu)^2]f(\varepsilon),
\label{kineq2}
\end{equation}
where the parameter
$\Upsilon$  is the only signature of the nonlinearity
of Eq.~(\ref{kineq1}).

Since Eq.~(\ref{kineq2}) represents an eigenvalue
problem for a linear operator, it can in general have more than
one solution. However, there is only one physical
solution, which satisfies the conditions
\begin{equation}
0\leq f(\varepsilon)\leq 1, \quad f(-\infty)=1, \quad f(+\infty)=0.
\label{conditions}
\end{equation}
Indeed, using a standard procedure one can show that two solutions
of the integral equation (\ref{kineq2}), $f_1$ and $f_2$,
corresponding to different parameters $\Upsilon_1\neq\Upsilon_2$
should be orthogonal, $\int d\varepsilon
f_1(\varepsilon)f_2(-\varepsilon)=0$. This contradicts the
conditions Eq.~(\ref{conditions}). The solution is also unique for the
same $\Upsilon$, i.e. it is not degenerate (for a proof, see
the Ref.\ \cite{SBL}). From Eq.~(\ref{kineq1}) and conditions
Eq.~(\ref{conditions}) it follows that if $f(\varepsilon)$ is a
solution then $1-f(-\varepsilon)$ also satisfies 
Eqs.~(\ref{kineq1}) and (\ref{conditions}). Since the solution is
unique, it has to have the symmetry
$f(\varepsilon)=1-f(-\varepsilon)$.

\begin{figure}[htb]
\centerline{\includegraphics[width=20pc]{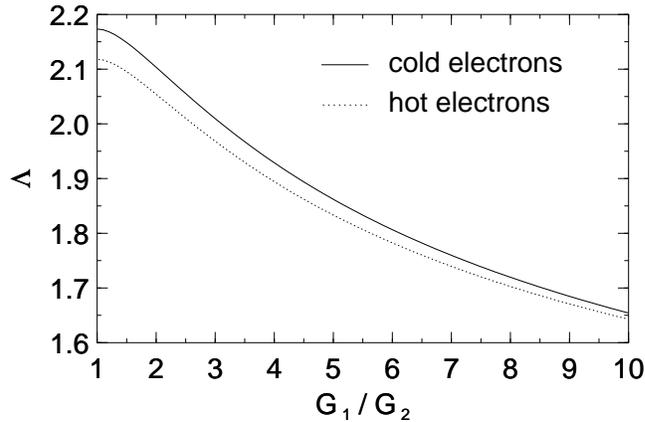}}
\caption{
The prefactor $\Lambda$ in the expression (\ref{I-continuum})
for the cotunneling current characterizes a universal cotunneling 
transport in the regime of weak cotunneling,
$\tau_{\rm in}\ll\tau_c$, ($\Lambda=1$, see 
Ref.\ \cite{averinazarov}),
and in the regime of strong cotunneling, $\tau_{\rm in}\gg\tau_c$
($\Lambda>1$). Here $\Lambda$ is
plotted as a function of $G_1/G_2$ (same as a function
of $G_2/G_1$) for the strong cotunneling,
for the cold-electron case, $\tau_{ee}\gg\tau_c$ (solid line)
and for the hot-electron case, $\tau_{ee}\ll\tau_c$ (dotted line).
$G_{1,2}$ are the tunneling conductances 
of a junctions connecting leads 1 and 2 with the QDS. 
}
\label{coldhot}
\end{figure}

We solve Eqs.~(\ref{kineq2}) and (\ref{conditions}) numerically 
and use Eqs.\ (\ref{Cpm}) and (\ref{F-continuum}) to find that 
the Fano factor is very close to 1 
(it does not exceed the value $F\approx 1.006$).
Next we use Eqs.\ (\ref{prefactor}) and (\ref{Upsilon})
to calculate the prefactor $\Lambda$ and plot the result as 
a function of the ratio of tunneling conductances, $G_1/G_2$,
(Fig.~\ref{coldhot}, solid line). For equal coupling
to the leads, $G_1=G_2$, the prefactor $\Lambda$ takes its maximum
value $2.173$, and thus the cotunneling current is approximately
twice as large compared to its value for the case of
weak cotunneling, $\tau_{\rm in}\ll\tau_c$. 
$\Lambda$ slowly decreases with increasing  
asymmetry of coupling and tends to its minimum value $\Lambda =1$ 
for the strongly asymmetric coupling
case $G_1/G_2$ or $G_2/G_1\gg 1$.

\subsection{Hot electrons}
\label{hot}

In the regime of hot electrons, $\tau_{ee}\ll\tau_c$, the
distribution is given by the equilibrium Fermi function
$f_F(\varepsilon)=\left[1+\exp(\varepsilon/k_BT_e)\right]^{-1}$, 
while the electron temperature $T_e$ has to be found
self-consistently from the kinetic equation.  
Eq.~(\ref{kineq1}) has to be modified to take into account 
electron-electron interactions. This can be done by adding the
electron collision integral $I_{ee}(\varepsilon)$ to the rhs
of (\ref{kineq1}).
Since the form of the distribution is known we need only the
energy balance equation, which can be derived by multiplying the
modified equation (\ref{kineq1}) by $\varepsilon$ and integrating
it over $\varepsilon$. The contribution from the collision
integral $I_{ee}(\varepsilon)$ vanishes, because the
electron-electron scattering conserves the energy of the system.
Using the symmetry $f_F(\varepsilon)=1-f_F(-\varepsilon)$ we
arrive at the following equation
\begin{equation}
\int\!\!\int d\varepsilon d\varepsilon' f_F(\varepsilon')[1-f_F(\varepsilon)]
\sigma(\varepsilon'-\varepsilon)\varepsilon =0.
\label{kineq3}
\end{equation}
Next we regroup the terms in this equation such that it
contains only integrals of the form $\int_{0}^{\infty}d\varepsilon
f_F(\varepsilon)(\ldots)$. This allows us to get rid of nonlinear
terms, and we arrive at the following equation,
\begin{equation}
\int d\varepsilon \varepsilon^3
[f_F(\varepsilon)-\theta(-\varepsilon)]
+3\Upsilon^2=\frac{(\Delta\mu)^4}{8(1+\lambda)}\,,
\label{kineq4}
\end{equation}
which holds also for the regime of cold electrons. Finally, we
calculate the integral in Eq.~(\ref{kineq4}) and express the result in
terms of the dimensionless parameter $\alpha=\Delta\mu/k_BT_e$,
\begin{equation}
\alpha=\pi\left[8(1+\lambda)/5\right]^{1/4}.
\label{alpha}
\end{equation}
Thus, since the distribution again depends on the ratio
$\varepsilon/\Delta\mu$, the hot electron regime is also
universal.

The next step is to substitute the Fermi distribution 
function with the temperature
given by Eq.~(\ref{alpha}) into Eq.~(\ref{Cpm}). 
We calculate the integrals and
arrive at the closed analytical expressions for the values of interest,
\begin{eqnarray}
&&
\Lambda=1+\frac{2\pi^2}{\alpha^2}=1+\sqrt{\frac{5}{2(1+\lambda)}}\,,
\label{I-hot}
\\ && F=1+\frac{12}{2\pi^2+\alpha^2}
\sum_{n=1}^{\infty}\left[\frac{1}{n^2} +\frac{2}{\alpha
n^3}\right]e^{-\alpha n}, \label{F-hot}
\end{eqnarray}
where again $\lambda=(G_1^2+G_2^2)/2G_1G_2\geq 1$.
It turns out that similar to the case of cold electrons,
Sec.\ \ref{cold}, the Fano factor for hot electrons is very 
close to $1$ (namely, it does not exceed the value $F\approx 1.007$). 
Therefore, we do not expect that the
super-Poissonian noise considered in this section
(i.e.\ the one which is due to heating 
of a large QDS caused by inelastic cotunneling through it) 
will be easy to observe in experiments. 
On the other hand, the transport-induced heating of a large 
QDS can be observed in the cotunneling current 
through the prefactor $\Lambda$,
which according to Eq.\ (\ref{I-hot}) takes its maximum value
$\Lambda= 1+\sqrt{5/4}\approx 2.118$ for $G_1=G_2$ and 
slowly reaches its minimum value $1$ with increasing (or decreasing) 
the ratio $G_1/G_2$ (see Fig.~\ref{coldhot}, dotted line).
Surprisingly, the two curves of $\Lambda$ vs $G_1/G_2$ for the
cold- and hot-electron regimes lie very close, 
which means that the effect of the electron-electron scattering
on the cotunneling transport is rather weak.

\section{Conclusions}

Here we give a short summary of our results.
In Sec.~\ref{FDT-Double}, we have derived the non-equilibrium FDT, i.e.\ the
universal relation (\ref{DB-FDT}) between the
current and the noise, for QDS in the weak cotunneling regime. Taking the limit $T,\omega\rightarrow 0$, we
show that the noise is Poissonian, i.e.\ $F=1$. 

In Sec.~\ref{Super-Poissonian}, we present the results of the 
microscopic theory of strong cotunneling, Ref.~\cite{SBL}: The master equation, Eq.~(\ref{MasterEq}), the average current, Eq.~(\ref{AvCurrent}), 
and the current correlators, Eqs.\ (\ref{NDnoise}) and (\ref{correlator-delta-s}),
for a QDS system coupled to leads
in the strong cotunneling regime $w_{\rm in}\ll I/e$ at small frequencies,
$\omega\ll \Delta_{\pm}$.
In contrast to sequential tunneling, where shot noise
is either Poissonian ($F=1$) or suppressed due to charge
conservation ($F<1$), we find that the noise in the
inelastic cotunneling regime can be super-Poissonian ($F>1$),
with a correction being as large as the Poissonian noise itself.
In the regime of elastic cotunneling $F=1$.

While the amount of super-Poissonian noise is merely estimated at the
end of Sec.~\ref{Super-Poissonian}, the noise of the cotunneling current
is calculated for the special case of a QDS with nearly degenerate states,
i.e.\ $\delta\! E_{nm}\ll \delta E$, in Sec.~\ref{degenerate}, where we
apply our results from  Sec.~\ref{Super-Poissonian}. The general solution
Eq.~(\ref{result01}) is further analyzed for two nearly degenerate
levels, with the result Eq.~(\ref{two-noise}).  More
information is gained in the specific case of a DD coupled to leads
considered in Sec.\ \ref{DD-system},
where we determine the average current Eqs.~(\ref{current4}-\ref{factor1}) 
and noise Eqs.~(\ref{noise2}-\ref{noise3}) 
in the weak cotunneling regime and the correlation correction to noise 
Eq.~(\ref{DD-noise}) in the strong cotunneling regime
as a function of frequency, bias, and the Aharonov-Bohm phase threading
the tunneling loop, finding signatures of the Aharonov-Bohm effect 
and of the quantum entanglement.

Finally, in Sec.~\ref{continuum}, another important situation is studied
in detail, the cotunneling through a QDS with a continuous energy spectrum,
$\delta E\ll\Delta\mu\ll E_C$.  Here, the correlation between tunneling
events plays a minor role as a source of super-Poissonian noise,
which is now caused by heating effects opening the possibility for
tunneling events in the reverse direction and thus to an enhanced
noise power.  In Eq.~(\ref{F-continuum}), we express the Fano factor $F$
in the continuum case in terms of the dimensionless numbers $C_\pm$,
defined in Eq.~(\ref{Cpm}), which depend on the electronic 
distribution function $f(\varepsilon)$ in the QDS (in this regime, a
description on the single-electron level is appropriate).
The current Eq.~(\ref{I-continuum}) is expressed in terms 
of the prefactor $\Lambda$, Eq.~(\ref{prefactor}).
Both $F$ and $\Lambda$ are then calculated for different regimes.
For weak cotunneling,
we immediately find $F=1$, as anticipated earlier, while for 
strong cotunneling we distinguish the two regimes of
cold ($\tau_{ee}\gg\tau_c$) and hot ($\tau_{ee}\ll\tau_c$) electrons.
For both regimes we find that the Fano factor is very close
to one, while $\Lambda$ is given in Fig.~\ref{coldhot}.

\begin{acknowledgements}
This work has been partially supported by the Swiss National
Science Foundation.
\end{acknowledgements}

\section*{Appendix A}

In this Appendix we present the derivation of Eqs.~\ref{DB-current}
and \ref{DB-noise}.
In order to simplify the intermediate steps, we use the
notation $\bar O(t)\equiv\int_{-\infty}^{t}dt'O(t')$ for any operator $O$,
and $O(0)\equiv O$.
We notice that, if an operator $O$ is a linear function of operators $D_l$ and $D_l^{\dag}$,
then $\bar O(\infty)=0$ (see the discussion in Sec.~\ref{FDT-Double}).
Next, the currents can be represented as the difference and the sum of $\hat I_1$
and $\hat I_2$,
\begin{eqnarray}
\hat I_d &=& (\hat I_2-\hat I_1)/2=ie(X^{\dag}-X)/2\,,
\label{A01}\\
\hat I_s &=& (\hat I_1+\hat I_2)/2=ie(Y^{\dag}-Y)/2\,,
\label{A02}
\end{eqnarray}
where $X=D_2+D_1^{\dag}$, and $Y=D_1+D_2$.
While for the perturbation we have
\begin{equation}
V=X+X^{\dag}=Y+Y^{\dag}\,.
\label{A03}
\end{equation}
First we concentrate on the derivation
of Eq.~(\ref{DB-current}) and redefine the average current 
Eq.~(\ref{current-noise}) as
$I=I_d$ (which gives the same result anyway, because the average number
of electrons on the QDS does not change $I_s =0$).

To proceed with our derivation, we make use of Eq.~(\ref{U-Operator}) and expand
the current up to fourth order in $T_{lkp}$:
\begin{equation}
I = i\int\limits^{0}_{-\infty}dt\int\limits^{t}_{-\infty}dt'
\langle \hat I_dV(t)V(t')\bar V(t')\rangle 
- i\int\limits^{0}_{-\infty}dt \langle \bar V\hat I_dV(t)\bar V(t)\rangle
+ {\rm c.c.}\,
\label{A04}
\end{equation}
Next, we use the cyclic property of trace to shift the time dependence to $\hat I_d$.
Then we complete the integral over time $t$ and use $\bar I_d(\infty)=0$. This procedure
allows us to combine first and second term in Eq.~(\ref{A04}),
\begin{equation}
I=-i\int\limits^{0}_{-\infty}dt
\langle [\bar I_dV+\bar V\hat I_d]V(t)\bar V(t)\rangle+{\rm c.c.}
\label{A05}\,
\end{equation}
Now, using Eqs.~(\ref{A01}) and~(\ref{A03})
we replace operators in Eq.~(\ref{A05}) with $X$ and $X^{\dag}$ in two steps:
$I=e\int^{0}_{-\infty}dt
\langle [\bar X^{\dag}X^{\dag}-\bar XX]V(t)\bar V(t)\rangle+{\rm c.c.}$,
where some terms cancel exactly. Then we work with $V(t)\bar V(t)$ and
notice that some terms cancel, because they are linear in $c_{lk}$ and
$c_{lk}^{\dag}$. Thus we obtain
$I = e\int^{0}_{-\infty}dt
\langle [\bar X^{\dag}X^{\dag}-\bar XX]
[X^{\dag}(t)\bar X^{\dag}(t)+X(t)\bar X(t)]\rangle
+{\rm c.c.}$.
Two terms $\bar XXX\bar X$ and
$\bar X^{\dag}X^{\dag}X^{\dag}\bar X^{\dag}$ describe tunneling
of two electrons from the same lead, and therefore they do not contribute
to the normal current. We then combine all other terms to extend the integral to $+\infty$,
\begin{equation}
I = e\int\limits^{\infty}_{-\infty}dt
\langle\bar X^{\dag}(t)X^{\dag}(t)X\bar X
-\bar XXX^{\dag}(t)\bar X^{\dag}(t)\rangle\,
\label{A07}
\end{equation}
Finally, we use
$\int^{\infty}_{-\infty}dt X(t)\bar X(t)=-\int^{\infty}_{-\infty}dt \bar X(t)X(t)$
(since $\bar X(\infty)=0$) to get Eq.~(\ref{DB-current}) with $A=X\bar X$.
Here, again, we drop
terms $D^{\dag}_1\bar D^{\dag}_1$ and $D_2\bar D_2$ responsible for
tunneling of two electrons from the same lead, and obtain $A$
as in Eq.~(\ref{DB-current}).

Next, we derive Eq.~(\ref{DB-noise}) for the noise power.
At small frequencies $\omega\ll \Delta_{\pm}$ fluctuations
of $I_s$ are suppressed because of charge conservation (see below),
and we can replace $\hat I_2$ in the correlator Eq.~(\ref{current-noise}) 
with $\hat I_d$.
We expand $S({\omega})$ up to fourth order in $T_{lkp}$,
use $\int_{-\infty}^{+\infty}dt\, \hat I_d(t)e^{\pm i\omega t}=0$,
and repeat the steps leading to Eq.~(\ref{A05}). Doing this we
obtain,
\begin{equation}
S(\omega)=-\int\limits^{\infty}_{-\infty}dt \cos(\omega t)
\langle [\bar V(t),\hat I_d(t)][\bar V, \hat I_d]\rangle\,.
\label{A08}
\end{equation}
Then, we replace $V$ and $\hat I_d$ with $X$ and $X^{\dag}$.
We again keep only terms relevant
for cotunneling,
and in addition we neglect terms of order $\omega/\Delta_{\pm}$
(applying same arguments as before, see Eq.~(\ref{A09})).
We then arrive at Eq.~(\ref{DB-noise})
with the operator $A$ given by Eq.~(\ref{DB-current}).

Finally, in order to show that fluctuations of $I_s$ are suppressed,
we replace $\hat I_d$ in Eq.~(\ref{A08}) with $\hat I_s$,
and then use the operators
$Y$ and $Y^{\dag}$ instead of $X$ and $X^{\dag}$.
In contrast to Eq.~(\ref{A07}) terms such as
$\bar Y^{\dag}Y^{\dag}Y\bar Y$ do not contribute,
because they contain integrals  of the form
$\int^{\infty}_{-\infty}dt\cos(\omega t) D_{l}(t)\bar D_{l'}(t)=0$.
The only nonzero contribution can be written as
\begin{equation}
S_{ss}(\omega)=\frac{e^2\omega^2}{4}\int\limits^{\infty}_{-\infty}dt \cos(\omega t)
\langle [\bar Y^{\dag}(t),\bar Y(t)][\bar Y^{\dag},\bar Y]\rangle\,,
\label{A09}
\end{equation}
where we have used integration by parts and the property $\bar Y(\infty)=0$.
Compared to Eq.~(\ref{DB-noise}) this expression contains an additional
integration over $t$, and thereby it is of order
$(\omega/\Delta_{\pm})^2$.

\section*{Appendix B}

In this Appendix we calculate the transition rates Eq.~(\ref{golden-rule}) for 
a DD coupled to leads with the coupling described by Eqs.~(\ref{DD-D})
and (\ref{equal}) and show that the four-level system in the singlet-triplet basis
Eq.~(\ref{basis})
can be mapped to a two-level system. For the moment we assume that the indices
$n$ and $m$ enumerate
the singlet-triplet basis, $n,m = S,T_0,T_{+},T_{-}$. Close to the sequential tunneling
peak, $\Delta_{-}\ll\Delta_{+}$, we keep only terms of the form $D^{\dag}_{l}\bar D_{l'}$.
Calculating the trace over the leads explicitly, we obtain at $T=0$,
\begin{eqnarray}
w_{nm}(l',l)
&=&
\frac{\pi\nu^2}{2\Delta^2_{-}}\,
\Theta(\mu_l-\mu_{l'}-\delta\! E_{nm})
\nonumber \\
& \times &
\sum_{j,j'}T^{*}_{lj}T_{lj'}T^{*}_{l'j'}T_{l'j}
M_{nm}(j,j')\,,
\label{C1}\\
M_{nm}(j,j')
&=& \sum_{s,s'}\langle n |d^{\dag}_{sj}d_{s'j}|m\rangle
\langle m |d^{\dag}_{s'j'}d_{sj'}|n\rangle\,,
\label{C2}
\end{eqnarray}
with 
$\Theta(\varepsilon)=\varepsilon \theta(\varepsilon)$,
and $\delta\! E_{nm}=0, \pm J$, and we have assumed $t_d\ll \Delta_{-}$.

Since the quantum dots are the same we get $M_{nm}(1,1)=M_{nm}(2,2)$,
and $M_{nm}(1,2)=M_{nm}(2,1)$. 
We calculate these matrix elements in the singlet-triplet basis
explicitly,
\begin{eqnarray}
&& M(1,1)=\frac{1}{2}\left(
\begin{array}{rrrr}
 1&1&1&1\\
1&1&1&1\\
1&1&2&0\\
1&1&0&2
\end{array}
\right),  
\label{C3} \\
&&M(1,2)=\frac{1}{2}\left(
\begin{array}{rrrr}
1&-1&-1&-1\\
-1&1&1&1\\
-1&1&2&0\\
-1&1&0&2
\end{array}
\right).  
\label{C4}
\end{eqnarray}
Assuming now equal coupling of the form Eq.~(\ref{equal})
we find that for $l=l'$ the matrix elements of the singlet-triplet
transition vanish (as we have expected, see Sec.\ \ref{degenerate}).
On the other hand the triplets are degenerate, i.e.\ $\delta\! E_{nm}=0$
in the triplet sector. Then from Eq.~(\ref{C1}) it follows
that $w_{nm}(l,l)=0$.
Next, we have $\Theta(\mu_2-\mu_{1}-\delta\! E_{nm})=0$,
since for nearly degenerate states we assume $\Delta\mu>|\delta\! E_{nm}|$,
and thus $w_{nm}(1,2)=0$.
Finally, for $w_{nm}=w^I_{nm}=w_{nm}(2,1)$ we obtain,
\begin{eqnarray}
w_{SS}
&=&
\frac{\pi}{2}\left(\frac{\nu{\cal T}^2}{\Delta_{-}}\right)^2     
\Delta\mu(1+\cos\phi),
\label{C5-1} \\
w_{ST}
&=&
\frac{\pi}{2}\left(\frac{\nu{\cal T}^2}{\Delta_{-}}\right)^2     
(\Delta\mu+J)(1-\cos\phi),
\label{C5-2}\\
w_{TS}
&=&
\frac{\pi}{2}\left(\frac{\nu{\cal T}^2}{\Delta_{-}}\right)^2     
(\Delta\mu-J)(1-\cos\phi),
\label{C5-3}\\
w_{TT}
&=&
\frac{\pi}{2}\left(\frac{\nu{\cal T}^2}{\Delta_{-}}\right)^2     
\Delta\mu
\nonumber\\
&\times&
\left(
\begin{array}{lll}
1+\cos\phi & 1+\cos\phi & 1+\cos\phi \\
1+\cos\phi & 2+2\cos\phi & 0\\ 
1+\cos\phi & 0 & 2+2\cos\phi  
\end{array}
\right).
\label{C5-4}
\end{eqnarray}

Next we prove the mapping to a two-level system. 
First we notice that because the matrix $w_{TT}$ is symmetric, 
the detailed balance equation for the stationary state gives 
$\bar\rho_n/\bar\rho_m=w_{mn}/w_{nm}=1$, $n,m\in T$.
Thus we can set $\bar\rho_{n}\to\bar\rho_2/3$, for $n\in T$.
The specific form of the transition matrix Eqs.~(\ref{C5-1}-\ref{C5-4})
helps us to complete the mapping by setting 
$(1/3)\sum_{m=2}^4w_{1m}\to w_{12}$,
$\sum_{n=2}^4w_{n1}\to w_{21}$, and
$(1/3)\sum_{n,m=2}^4w_{nm}\to w_{22}$, so that 
we get the new transition matrix Eq.~(\ref{DD-rates}), while 
the stationary master equation for the new two-level density matrix
does not change its form. If in addition we set 
$(1/3)\sum^{4}_{m=2}\delta\rho_{1m}(t)\to\delta\rho_{12}(t)$,
$\sum^4_{n=2}\delta\rho_{n1}(t)\to\delta\rho_{21}(t)$, and
$(1/3)\sum^{4}_{n,m=2}\delta\rho_{nm}(t)\to\delta\rho_{22}(t)$,
then the master equation Eq.~(\ref{MasterEq}) for $\delta\rho_{nm}(t)$ and
the initial condition $\delta\rho_{nm}(0)=\delta_{nm}-\bar\rho_{n}$
do not change either. Finally, one can see that under this mapping
Eq.~(\ref{correlator-delta-s}) 
for the correction to the noise power $\Delta S(\omega)$
remains unchanged. Thus we have accomplished 
the mapping of our singlet-triplet system 
to the two-level system with the new transition 
matrix given by Eq.~(\ref{DD-rates}).

\end{article}                            

\begin{thebibliography}{99}
\bibitem{AB} The Aharonov-Bohm effect in the noise of
non-interacting electrons in a mesoscopic ring has been discussed
in Ref.\ \cite{review}.
\bibitem{comment5}
One could view this as an analog of a whistle effect,
where the flow of air (current) is strongly modulated by a bistable state
in the whistle, and vice versa. The analogy, however, is not
complete,
since the current through the QDS is random due to quantum fluctuations.
\bibitem{inelastic}
Note that the Aharonov-Bohm effect is not suppressed by this inelastic cotunneling,
since the {\it entire} cotunneling process involving also leads is elastic:
the initial and final states of the {\it entire} system have the same energy.
\end{thebibliography}
\end{document}